\documentclass[aps,prb,twocolumn,showpacs,amsmath,amssymb,floatfix]{revtex4}
\usepackage{epsfig}
\usepackage{bm}
\def\l{\langle\!\langle}
\def\r{\rangle\!\rangle}

\begin{document}
\title{Stochastic dynamics of a Josephson junction threshold detector}
\author{Eugene V. Sukhorukov$^1$ and Andrew N. Jordan$^2$}
\affiliation{$^1$\
D\'{e}partment de Physique Th\'{e}orique, Universit\'{e} de Gen\`{e}ve,
CH-1211 Gen\`{e}ve 4, Switzerland\\
$^2$\ Department of Physics and Astronomy, University of Rochester, Rochester, New York 14627, USA}

\date{\today}

\begin{abstract}
We generalize the stochastic path integral formalism by considering Hamiltonian dynamics 
in the presence of general Markovian noise. Kramers' solution of the activation rate for 
escape over a barrier is generalized for non-Gaussian driving noise in both the overdamped 
and underdamped limit. We apply our general results to a Josephson junction detector 
measuring the electron counting statistics of a mesoscopic conductor.  Activation rate dependence 
on the third current cumulant includes an additional term originating from the
back-action of the measurement circuit.
\end{abstract}


\maketitle

\section{Introduction}

Detecting electron counting statistics has become a major experimental challenge in 
mesoscopic physics.
First attempts to measure non-Gaussian effects in current noise have revealed that the detection
problem is quite subtle.  In particular, the experiment\cite{Reulet} 
found that the third current cumulant was not described by the simple theoretical prediction,\cite{Levitov1} but was masked by the influence of the measurement circuit causing an additional ``cascade'' correction.\cite{Beenakker,Nagaev}  Recent experiments demonstrated a
measurement of the third current cumulant without cascade corrections,\cite{Bomze} and the detection of individual electron counting statistics.\cite{Simon}  Stringent bandwidth requirements in measuring the third cumulant suggested that further experimental advances would require a new approach.

\begin{figure}[htb]
\epsfxsize=6.5cm
\epsfbox{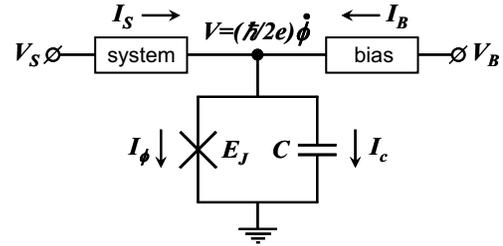}
\caption{Josephson junction (JJ) threshold detector: Simplified electrical circuit 
with the JJ (marked with an X) and mesoscopic system inserted.  
The driving noise of the system creates fluctuations of the center node voltage $V$, 
that can activate the JJ out of its supercurrent state, into its running dissipative 
state until it is recaptured.
}
\vspace{-5mm}
\label{circuit}
\end{figure}

A conceptually different way to measure rare current fluctuations is with a threshold 
detector,\cite{Nazarov,ourPRB} the basic idea of which is analogous to a pole vault:  A detection event occurs when the measured system variable exceeds a given value.  A natural candidate for such a detector is a metastable system operating on an activation 
principle.\cite{Kramers} By measuring the rate of switching out of the metastable state, information about the statistical properties of the noise driving the system may be extracted.  A threshold detector using an on-chip conductor which contains a region of negative differential resistance\cite{ourprl,ourPRB} was proposed by the authors and shown to be capable of measuring large deviations of current. Tobiska and Nazarov proposed a Josephson junction (JJ) threshold detector,\cite{Nazarov} the simplest variant of which 
(see Fig.\ \ref{circuit}) operates essentially in a Gaussian regime.\cite{Pekola}  
The third cumulant contribution is small\cite{Ankerhold} and may be extracted using the asymmetry of the switching rate with respect to bias current, as has been demonstrated in recent experiments.\cite{Pekola2,Saclay}
  
In this Letter we solve Kramers' problem\cite{Kramers} 
of noise-activated escape from a metastable 
state beyond the Gaussian noise approximation and investigate how the measurement circuit 
affects threshold detection. Starting with general Hamiltonian-Langevin 
equations which includes deterministic dynamics, dissipation, and fluctuations, we represent 
the solution as a stochastic path integral of Hamiltonian form\cite{SPI,JMP} by doubling 
the number of degrees of freedom. In the weak damping case, the dynamics is dominated by energy 
diffusion, which we account for by a change of variables, enabling an effectively two-dimensional 
representation. We calculate the escape rate via an instanton 
calculation, and obtain a formal solution of Kramers' problem.\cite{Kramers} 
Applying these general results to a JJ threshold detector, we account for the 
influence of the measurement circuit and find that the cascade corrections
are a consequence of the non-equilibrium character of the noise.\cite{footnote0} 

\section{Hamiltonian-Langevin equations}

In order to reformulate and solve Kramers' problem beyond Gaussian noise, we introduce a theoretical framework that relies on a separation of time scales: Slow motion of a deterministic system on a time scale $T_0$ is affected by quickly fluctuating noise sources with correlation time $\tau_0 \ll T_0$. Quite generally, this situation can be described by Hamiltonian-Langevin (HL) equations. For instance, in a two-dimensional phase space $(p,q)$ the equations of motion are
\begin{equation}
\dot q=\partial H(p,q)/\partial p+I_q,\quad
\dot p=-\partial H(p,q)/\partial q+I_p,
\label{HLE}
\end{equation}
where $H(p,q)$ is the Hamiltonian that generates deterministic motion, and $I_q$ and $I_p$ are white noise 
sources. On the intermediate time scale $t$, such that $T_0\gg t\gg\tau_0$, the sources are Markovian and fully characterized by the generating function ${\cal H}(\lambda_p,\lambda_q)$ of their cumulants (irreducible moments): 
$\l I_p^nI_q^m\r=\partial^n_{\lambda_p}\partial^m_{\lambda_q}{\cal H}(0,0)$.

While Langevin equations may be solved by standard methods in stochastic physics,\cite{book} the methods fail when applied to higher-order cumulants.  The reason is that slow variations of $p$ and $q$ affect the noise sources and lead to {\em cascade corrections},\cite{Nagaev,JMP} which cannot be obtained from linearized differential equations (\ref{HLE}).  Following the steps of Refs.\ [\onlinecite{JMP,SPI}], we represent the slow evolution of the system by the stochastic path integral (SPI) $P=\int\! {\cal D}\Lambda\!\int\!{\cal D}{\bf R}\exp(S)$, where the action $S$ is given in the explicitly canonically invariant form as
\begin{equation}
S=\int dt'[-\Lambda\cdot \dot {\bf R}+\Lambda\cdot \{{\bf R},H\}+
{\cal H}(\Lambda, {\bf R})].
\label{action}
\end{equation}
Here ${\bf R}=(p,q)$ and $\Lambda=(\lambda_p,\lambda_q)$ are the sets of physical
and canonically conjugated ``counting'' variables, respectively, and $\{\ldots\}$
denotes the Poisson bracket with respect to $p$ and $q$.
By fixing ${\bf R}$ in the final state of (\ref{action}) we obtain the 
probability distribution $P({\bf R})$, while fixing the final $\Lambda$ 
variables turns the SPI into the moment generating function $P(\Lambda)$.

The large parameter $T_0/\tau_0\gg 1$ allows the saddle-point evaluation of the SPI
and thus requires solving Hamilton's equations of motion in the extended space.\cite{JMP} 
There always exists a trivial solution $\Lambda=0$ and $\dot {\bf R}=\{{\bf R},H\}+\langle {\bf I}\rangle$, where ${\bf I}\equiv(I_p,I_q)$, that describes the ``average'' dynamics in physical space with a null action $S=0$, giving the proper normalization of the distribution $P$. In the context of the generalized Kramers' problem, one has to find a non-trivial {\em instanton} solution $\Lambda_{\rm in}({\bf R})$ of Eq.\ (\ref{action}), for which the SPI gives a rate of the noise activated escape from a localized state, $\Gamma\propto\exp(S_{\rm in})$. \cite{footnote1}

\section{Energy diffusion and escape rate}

As an important example, let us consider quasi-periodic motion with the period $T_0\gg\tau_0$. We change variables to energy $E=H(p,q)$ and ``time'' $s=s(p,q)$,\cite{footnote2}
accompanied by a new set of counting variables $(\lambda_E, \lambda_s)$,
which are defined via $\lambda_p=\lambda_E\,\partial_p H+\lambda_s\partial_p s$
and $\lambda_q=\lambda_E\,\partial_q H+\lambda_s\partial_q s$.  
Using $\{E,H\}=0$ and $\{s,H\}=1$, we obtain $\Lambda\cdot \{{\bf R},H\}=\lambda_s$. 
Fluctuations of the $s$ variable increase the action without leading to escape, 
therefore they can be neglected
by choosing $\lambda_s=0$. In the weak damping regime, one can set $s=t$, so that $\lambda_p=\dot q\lambda_E$ and $\lambda_q=-\dot p\lambda_E$. The action for the energy diffusion then reads, 
\begin{equation}
S=\int dt'[-\lambda_E\dot E+{\cal H}(\lambda_E\dot q,-\lambda_E\dot p)].
\label{diffusion}
\end{equation} 

The following intuitive argument supports this result. The energy balance equation $\dot E=\dot qI_p-\dot p I_q$ follows directly from the HL equations (\ref{HLE}) and takes the form of a Langevin equation.  If $T_0\gg\tau_0$, then variables $\dot q$ and $\dot p$ change slowly and can be considered as being ``effective charges'', which after following [\onlinecite{JMP}]
leads to Eq.\ (\ref{diffusion}).

To leading order in weak damping we can replace the generator ${\cal H}$ in Eq.\ (\ref{diffusion})
with its average over the oscillation period $\langle{\cal H}\rangle_E\equiv(1/T_0)\oint dt {\cal H}$, 
evaluated for fixed $\lambda_E$ and $E$.  Corrections in damping may be found by 
taking into account slow energy dissipation $\dot E=\partial_{\lambda_E}{\cal H}$ and  
$\dot \lambda_E=-\partial_E{\cal H}$, 
while averaging over the period $T_0$.
We are interested in the instanton solution $\lambda_E=\lambda_{\rm in}(E)$ with $\langle{\cal H}\rangle_E=0$ 
in the initial and final state.\cite{ourprl}  Since the ``Hamiltonian'' $\langle{\cal H}\rangle_E$ is 
an integral of motion, we obtain an important result for the escape rate,
\begin{equation}
\log\Gamma=-\int \lambda_{\rm in} dE,\quad \langle
{\cal H}(\lambda_{\rm in}\dot q,-\lambda_{\rm in}\dot p)\rangle_E=0,
\label{instanton}
\end{equation}
which formally solves Kramers' problem for arbitrary Markovian noise in the weak damping limit. 
Below we apply the theory outlined here to the stochastic dynamics of a Josephson threshold detector.

\section{Josephson threshold detector}

The circuit in Fig. 1 shows the essential part of the detector comprised of the JJ with Josephson energy $E_J$, and the capacitor, $C$. The circuit is current-biased with $I_B$ through the macroscopic conductor and by the system current $I_S$, which is to be measured. According to Kirchhoff's law, the total bias current $I_S+I_B$ is equal to the sum of the Josephson current $I_{\phi}=(E_J/\Phi_0)\sin\phi$ where $\Phi_0=\hbar/2e$, and the displacement current $I_C=C\dot V$. This leads to the equation of motion for the superconducting phase $\phi$, 
\begin{equation}
C\Phi_0^2\ddot\phi+E_J\sin\phi=\Phi_0(I_S+I_B),
\label{motion}
\end{equation}
where we used the relation $V=\Phi_0\dot\phi$. 

In order to simplify the following analysis and concentrate on our main message, we make a number of assumptions, most of which will be relaxed later. First, we consider an ohmic system and bias resistor, so that $\langle I_S\rangle=J_S-G_SV$ and $\langle I_B\rangle=J_B-G_BV$, where $G_S$, $G_B$ are the system and bias conductances, and the constant currents $J_S=G_SV_S$, $J_B=G_BV_B$ are just tunable parameters.   The bias resistor, being a macroscopic system, creates Gaussian Nyquist noise $\l I^2_B\r=2TG_B$.  We further assume a high-impedance circuit, so that the back flow part of the bias current $G_BV$ and the Nyquist noise may be neglected, $I_B=J_B$, which we refer to as the {\em ideal detection} scheme. Then Eq.\ (\ref{motion}) can be rewritten as a set of HL equations (\ref{HLE}) for the phase variable $\phi$ and canonically conjugated momentum $p=\Phi_0 Q$ (where $Q=CV$ is the total charge on the capacitor)
\begin{equation}
\dot\phi=p/m,\quad \dot p=-\partial U/\partial\phi + \Phi_0(I_S-J_S),
\label{HLE-phase}
\end{equation}
with ``mass'' $m=\Phi_0^2C$.  

Equations (\ref{HLE-phase}) describe the motion of a ``particle'' in the tilted periodic potential
\begin{equation}
U(\phi)=-E_J\cos\phi-\Phi_0(J_S+J_B)\phi,
\label{potential}
\end{equation}
stimulated by the dissipative part $I_S-J_S$ of the system's current.  For later convenience, we define the normalized total bias current as ${\cal J}=\Phi_0(J_S+J_B)/E_J$.  Dissipation leads to relaxation of the JJ into one of its metastable states.  In the supercurrent state, the phase is localized in one of the potential wells, so that $\langle V\rangle=0$. In the dissipative state, the phase drifts along the bias which generates the non-zero voltage drop $V$.   
In the remaining part of the paper we investigate noise activated escape from the localized state.

\section{Weak damping regime}

Here the system conductance is small, $G_S\ll \omega_{\rm pl}C$, so the phase oscillates with the plasma frequency $\omega_{\rm pl}=\Omega_J(1-{\cal J})^{1/4}$, where $\Omega_J=(E_J/\Phi_0^2C)^{1/2}$.  The energy relaxes slowly with the rate $G_S/C$ to the local potential minimum. We further assume the separation of time scales, 
$1/\tau_0\sim\max\{eV_S, T\}\gg\hbar\omega_{\rm pl}$, so that the noise source $I_S$ is Markovian. 
Comparing (\ref{HLE-phase}) and (\ref{HLE}), we identify $q=\phi$, $I_p=\Phi_0(I_S-J_S)$ and $I_q=0$, 
so the equations for the escape rate and ``instanton line'' read
\begin{equation}
\log\Gamma=-\Phi_0^{-1}\int \lambda_{\rm in} dE, 
\quad \langle{\cal H}(\lambda_{\rm in}\dot\phi)\rangle_E=0,
\label{WDR-instanton}
\end{equation}
where ${\cal H}$ is the generator of the cumulants of the dissipative part of the system current, $I_S-J_S$, and $\Phi_0$ plays the role of an effective charge. 

\begin{figure}[t]
\epsfxsize=7cm
\epsfbox{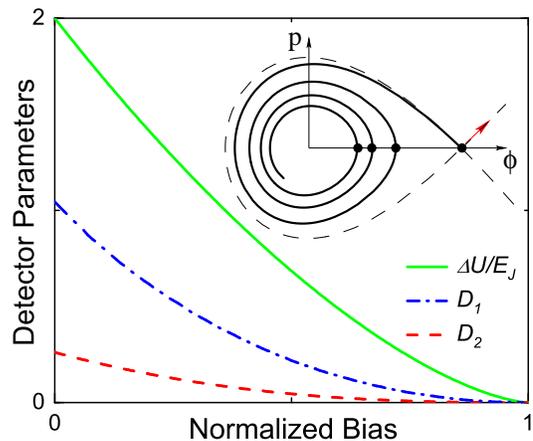}
\caption{The normalized potential threshold $\Delta U/E_J$
and dimensionless factors  $D_1$ and $D_2$, are plotted versus the normalized 
current bias ${\cal J}=\Phi_0(J_S+J_B)/E_J$. Inset shows the phase space plot
of a trajectory for weak damping (thick line) which leads to the escape, and the 
separatrix (dashed line). The turning points $\phi_0$ are shown by the dots. 
}
\vspace{-5mm}
\label{parameters}
\end{figure}

We first consider the system in thermal equilibrium with Gaussian Nyquist noise, $\l I_S^2\r = 2TG_S$, 
as a checkpoint of the theory. The cumulant generator acquires the simple form ${\cal H}=-\Phi_0G_S\dot\phi(\lambda_{\rm in}\dot\phi)+ TG_S(\lambda_{\rm in}\dot\phi)^2$, where the first term comes from the linear response of the system current $I_S-J_S=-G_SV$ to the potential $V=\Phi_0\dot\phi$, while the second term is the noise contribution.  Averaging ${\cal H}$ over the period of oscillations, we observe that the term $\langle\dot\phi^2\rangle_E$ cancels, so that $\lambda_{\rm in}=\Phi_0/T$. Using Eq.\ (\ref{WDR-instanton}) we obtain Kramers' well-known formula for the rate of thermally activated escape
\begin{equation}
\log\Gamma=-\Delta U/T,
\label{Kramers}
\end{equation}  
where the potential threshold $\Delta U$ is a function of normalized bias ${\cal J}$. For the potential (\ref{potential}) one obtains $\Delta U/E_J=2(1-{\cal J}^2)^{1/2}- 2 {\cal J}\arccos{\cal J}$, 
see Fig.\ \ref{parameters}.   Limiting values are $\Delta U/E_J=2$ for ${\cal J}=0$, 
while as ${\cal J} \rightarrow 1$,  
$\Delta U/E_J \approx (4\sqrt{2}/3) (1-{\cal J})^{3/2}$.

The above example shows that the argument of ${\cal H}$ is small.  Indeed, we estimate 
$\dot\phi\sim\omega_{\rm pl}$, so $e\dot\phi\lambda_{\rm in}\sim\hbar\omega_{\rm pl}/T\ll 1$ due to the separation of time scales. Therefore, for any system away from equilibrium we can expand $\langle {\cal H}\rangle_E$ in Eq.\ (\ref{WDR-instanton}), average the result over the period of oscillations, and invert the series for $\lambda_{\rm in}$. To second order in $\lambda_{\rm in}$ we again obtain Kramers' formula (\ref{Kramers}) with the temperature $T$ replaced with the effective temperature 
\begin{equation}
T_{\rm eff}\equiv \l I_S^2\r/2\,G_S
\label{temperature}
\end{equation} 
of the non-equilibrium noise. 

To find the third cumulant correction to the Kramers' formula, we expand $\langle {\cal H}({\dot \phi} \lambda_{\rm in})\rangle_E$ in $\lambda_{\rm in}$ and the voltage $V$, and collect all terms to next order in small parameter $\hbar\omega_{\rm pl}/T_{\rm eff}$. Two terms which contain $\l I_S^3\r$ and $\partial_{V_S}\!\l I_S^2\r$ come with the factor $\langle\dot\phi^3\rangle_E$.\cite{footnote3} This factor vanishes to leading order in damping, because for fixed energy $\dot\phi$ is an odd function of time over the oscillation period. Therefore, we evaluate the average $\langle\dot\phi^3\rangle_E$ by taking into account slow variation of the energy. This is accomplished by taking functional variations of the
action to find $\lambda_{\rm in}=\Phi_0/T_{\rm eff}$ and 
$\dot E=\Phi_0^2G_S\dot\phi^2$. 
We skip a number of straightforward steps (see Appendix) and present the result for the 
non-equilibrium escape rate 
\begin{subequations}
\begin{eqnarray}
\log\Gamma &=& -\frac{\Delta U}{T_{\rm eff}}+
\frac{D_1\Phi_0E_J\l I_S^3\r_{\rm tot}}
{CT_{\rm eff}^3},
\label{rate1}\\
D_1 &=& \int \frac{dE}{E_J}\,
\frac{\langle (\phi_0-\phi)\dot\phi^2\rangle_E}
{3\langle\dot\phi^2\rangle_E}\,,
\label{D1}
\end{eqnarray}
\label{WDR}
\end{subequations}
where the dimensionless factor $D_1({\cal J})$ is a characteristic of the detector. Appearing in the integral (\ref{D1}) is one of the turning points of the oscillating phase, $\phi_0$, defined by the solution of $U(\phi_0)=E$ that is nearest the top of the potential.  
For the potential (\ref{potential}) $D_1(0)=\pi/3$ and for strong bias ${\cal J}\to 1$ we obtain $D_1({\cal J})\approx 0.8 (1-{\cal J})^2$.  The parameter $D_1$, evaluated numerically, is shown in Fig.\ \ref{parameters}.

The total third current cumulant
\begin{equation}
\l I_S^3\r_{\rm tot}=\l I_S^3\r-3T_{\rm eff}\,\partial_{V_S}\!\l I_S^2\r
\label{cumulant1}
\end{equation}
is taken at $V=0$ and contains a correction originating from the slowly varying second current cumulant. This contribution is analogous to the cascade correction\cite{Nagaev,Beenakker} directly observed in third cumulant.\cite{Reulet}  The correction is generally not small and may even change the sign of the total cumulant. 
For instance, for systems far from equilibrium $\l I_S^n\r=F_n\langle I_S\rangle$, where $F_n$ are the dimensionless Fano factors, we obtain $\l I_S^3\r_{\rm tot}=(F_3-3F_2^2/2)\langle I_S\rangle$.

\section{Overdamped regime}

In this regime the conductance is large, $G_S\gg \omega_{\rm pl}C$, and the dynamics is entirely due to slow phase relaxation with the rate $\omega_{\rm pl}^2C/G_S$. Therefore, we can 
set $p=m\dot\phi$ and $\lambda_q=0$, and neglect the first term in the action (\ref{action}), 
so that the action reads: $S=\int dt\,[{\cal H}(\Phi_0\lambda_p,\dot\phi)-\lambda_p\partial_\phi U]$, where again ${\cal H}$ is the generator of the cumulants of the dissipative part of the system current, $I_S-J_S$.

The following analysis is analogous to that of the weak damping regime. We first expand ${\cal H}$ to second order in $\lambda_p$ and find from the equations of motion that $\dot\phi=T_{\rm eff}\lambda_p$ and $\lambda_p=\partial_\phi U/(\Phi_0^2T_{\rm eff}G_S)$. Substituting these results back to the action we immediately obtain Kramers' formula (\ref{Kramers}) with $T$ replaced with $T_{\rm eff}$.  
Next we observe that the argument of ${\cal H}$ is small, namely 
$e\Phi_0\lambda_p\sim \hbar \omega_{\rm pl}^2C/G_ST_{\rm eff}\ll 1$ 
due to the separation of time scales requirement. 
We collect all the terms to third order in this parameter 
and evaluate them perturbatively using the above results for $\lambda_p$ and 
$\dot\phi$. We finally obtain
\begin{subequations}
\begin{eqnarray}
\log\Gamma &=& -\frac{\Delta U}{T_{\rm eff}}+
\frac{D_2E_J^2\,\l I_S^3\r_{\rm tot}}
{\Phi_0T_{\rm eff}^3\,G_S^2},
\label{rate3}\\
D_2 &=&(1/6)\int d\phi\,(\partial_\phi U/E_J)^2,
\label{D2}
\end{eqnarray}
\label{SDR}
\end{subequations}
where the total cumulant $\l I_S^3\r_{\rm tot}$ is given by Eq.\ (\ref{cumulant1}), and $D_2({\cal J})$ is a dimensionless detector property. For the potential (\ref{potential}) it is given by $D_2=(1/6)(1+2{\cal J}^2)\arccos{\cal J}- (1/2){\cal J}(1-{\cal J}^2)^{1/2}$, see Fig.\ 2.  Limits are $D_2(0) = \pi/12$, and as ${\cal J} \rightarrow 1$, $D_2({\cal J}) \approx  
0.25(1-{\cal J})^{5/2}$.

\section{Discussion}

We now remark on the application of our results to the detection of non-Gaussian fluctuations. It is evident from Fig.\ \ref{parameters} that $D_{1,2}\ll \Delta U/E_J$ as ${\cal J}\to 1$. Therefore, in the strong bias regime the third cumulant contribution in (\ref{WDR}) and (\ref{SDR}) is suppressed compared to the Kramers' term $S_0=\Delta U/T_{\rm eff}$. But even for a relatively weak bias, when $D_{1,2}\sim 1$, non-Gaussian effects are small. 
Indeed, it is easy to estimate the correction as $({\cal R}_{\rm wd}/{\cal Q})S_0$, 
where the JJ  quality factor ${\cal Q}=C\omega_{\rm pl}/G_S>1$ 
in the weak damping regime (wd), and the ratio 
${\cal R}_{\rm wd}=\hbar\omega_{\rm pl}/T_{\rm eff}< 1$ due to
the separation of time scales. Similarly, in the strong damping regime (sd), 
${\cal Q}<1$, the correction is of order ${\cal R}_{\rm sd}S_0$, 
where  ${\cal R}_{\rm sd}=\hbar \omega_{\rm pl}^2C/G_ST_{\rm eff}< 1$
due to the separation of time scales.
Since $S_0$ itself cannot be too large for the escape to be detected, in experiments one should try to saturate the above inequalities and use the asymmetry of the third cumulant (\ref{cumulant1}) as a function of the current bias $J_B$.

Next, we briefly discuss nonideal detection and nonlinear effects.  The finite conductance of the bias resistor $G_B$ contributes to the total conductance of the circuit $G_{\rm tot}=G_S+G_B$ and the Nyquist noise $\l I_B^2\r=2TG_B$ adds to the total noise power. Therefore, in Eqs.\ (\ref{WDR}-\ref{SDR}) one has to replace 
$G_S$ with $G_{\rm tot}$ and the effective temperature with 
$\tilde T=(G_ST_{\rm eff}+G_BT)/G_{\rm tot}$. 
The nonlinearity of the system current leads to the
correction $(\lambda_p/2)(\Phi_0\dot\phi)^2\partial_{V_S}^2\langle I_S\rangle$ to ${\cal H}$,
and thereby to an additional contribution to the total third cumulant, 
$\l I_S^3\r_{\rm tot}=\l I_S^3\r-3\tilde T\,\partial_{V_S}\!\l I_S^2\r
+3\tilde T^2\partial^2_{V_S}\langle I_S\rangle$, 
in both transport regimes considered above.  
Quite remarkably, this total third cumulant is related via 
$\l I_S^3\r_{\rm tot}=3C^2G_{\rm tot}\l V^3\r$ to the instantaneous fluctuations of the voltage $V$.\cite{footnote4}

We finally note that in the experiment of Ref.\ [\onlinecite{Pekola2}] the circuit 
corrections have not been observed, which we explain by a very low impedance of 
the circuit, $G_S/G_{\rm tot}\ll 1$, operating in the weak
damping regime with ${\cal Q}=C\omega_{\rm pl}/G_{\rm tot}=2.5$. Indeed, in this case 
$\tilde T=(G_S/2G_{\rm tot})\,eV_S$ (assuming $T=0$), therefore the circuit 
corrections are suppressed by a small factor $G_S/G_{\rm tot}$. 
On the other hand, the first term in Eq.\ (\ref{rate1}) is $S_0=\Delta U/\tilde T$ 
and the second term can be estimated as 
$(G_{\rm tot}/G_S)({\cal R}_{\rm wd}/{\cal Q})S_0$, 
where ${\cal R}_{\rm wd}=\hbar\omega_{\rm pl}/eV_S$ in this case.
Therefore the relative contribution of the third cumulant to the total action 
increases by the factor $G_{\rm tot}/G_S$, which makes it favorable to use a 
low-impedance circuit.

In conclusion, we have formally solved Kramers' problem of noise-activated 
escape from a metastable state for Markovian non-equilibrium noise beyond 
the Gaussian approximation. We have applied this result to the Josephson
junction threshold detector and evaluated the third current cumulant 
contribution to the escape rate. The back action of the electrical circuit 
on the measured noise is entirely a non-equilibrium effect, which leads 
to ``cascade'' corrections for the third cumulant. It would be interesting 
to apply our theory to the dynamics of the Josephson bifurcation 
amplifier.\cite{amplifier}

\acknowledgements
This work has been supported by the Swiss National
Science Foundation.

\appendix

\section{}

In this Appendix, we give the details of the derivation of Eq.\ (\ref{WDR}).  
Starting with Eq.\ (\ref{WDR-instanton}), 
we expand the generating function in the counting variable to third order,
\begin{equation}
{\cal H} = \langle I_S -J_S\rangle \lambda_{\rm in} {\dot \phi} + \l I^2_S\r (\lambda_{\rm in} {\dot \phi})^2/2 + \l I^3_S\r(\lambda_{\rm in} {\dot \phi})^3/3! +\ldots
\label{A1}
\end{equation}
Taking the assumptions in the text of an ohmic system conductor, we replace $\langle I_S-J_S\rangle = - G_S V$.  The noise produced by the system conductor depends on the dynamically changing voltage drop $V_S-V$ (see Fig.\ \ref{circuit}), so we also expand the second current cumulant in the voltage $V$, namely 
$ \l I^2_S\r =  \l I^2_S\r - \partial_{V_S} \l I^2_S\r V + \ldots$, where we recall that $V_S$ is the voltage across the system, and the coefficients above are evaluated at $V=0$. Using the Josephson relation $V = \Phi_0 {\dot \phi}$, and averaging (\ref{A1}) over the oscillation period at fixed energy, we now need to find the instanton line and solve the equation
\begin{eqnarray}
0 &=& - G_S \Phi_0  \langle {\dot \phi}^2\rangle_E  \lambda_{\rm in} + \l I^2_S\r \langle {\dot \phi}^2\rangle_E \lambda_{\rm in}^2/2 \\ &-&
\Phi_0 \partial_{V_S} \l I_S^2\r \langle {\dot \phi}^3\rangle_E \lambda_{\rm in}^2/2 +   \l I_S^3\r  \langle {\dot \phi}^3\rangle_E \lambda_{\rm in}^3/3! +\ldots \nonumber
\end{eqnarray}
for $\lambda_{\rm in}(E)$.  

As argued in the text, the instanton solution $\lambda_{\rm in}(E)$ is small for the JJ, so the series may be inverted to find to third order in $\hbar \omega_{\rm pl}/T_{\rm eff}$,
\begin{equation}
\lambda_{\rm in} = \frac{\Phi_0}{T_{\rm eff}} - \frac{ \l I_S^3\r_{\rm tot}  \Phi_0^2 \langle {\dot \phi}^3\rangle_E }{6 T_{\rm eff}^3 G_S \langle {\dot \phi}^2\rangle_E}.
\end{equation}
Inserting this instanton solution into the action, we obtain for the activation rate
\begin{equation}
\log \Gamma = 
-\frac{\Delta U}{T_{\rm eff}} 
+ \frac{\Phi_0 \l I_S^3\r_{\rm tot}}{6 T_{\rm eff}^3 G_S } 
\int dE \frac{\langle {\dot \phi}^3\rangle_E}
{\langle {\dot \phi}^2\rangle_E}\;.
\label{result}
\end{equation}

However, to leading order in damping, the average $\langle {\dot \phi}^3\rangle_E$ vanishes because ${\dot \phi}$ is an odd function of time over the interval of one period.  Therefore, we must evaluate this average to next order in the variation of energy 
over one period. Recalling that the action 
is $S = \int dt' [-\lambda_E {\dot E}/\Phi_0+ {\cal H}(\lambda_E\dot\phi) ]$, 
the time dependence of the energy $E$ may be 
found by taking the functional derivative 
$\delta S/\delta\lambda_E$  
to find ${\dot E} = \Phi_0 \partial_{\lambda_E} {\cal H}$.  
Inserting the instanton solution $\lambda_{\rm in} = \Phi_0/T_{\rm eff}$, we obtain 
\begin{equation}
{\dot E} = \Phi_0^2 G_S {\dot \phi}^2,
\label{edot}
\end{equation}
so the energy is increasing. 

At a given value of $\phi$, the linear correction 
to ${\dot \phi}^2$ due to energy dissipation is given 
by $\delta {\dot \phi}^2 = (2/m)\delta E$. Therefore,
we use $dt\dot\phi=d\phi$ to write 
\begin{equation}
\langle {\dot \phi}^3\rangle_E = 
\frac{1}{T_0} \oint d\phi {\dot \phi}^2=
\frac{2}{m T_0} \oint d\phi\,  \delta E,
\end{equation}
where the integral starts from the turning point nearest to the top of the barrier, 
$\phi = \phi_0$.  This is done because the escape of the JJ into the running 
state happens at the top of the barrier, so the last part of the trajectory before 
escape must be fully accounted for.  

Integrating by parts, $\oint d\phi\delta E = \phi \delta E\vert_{0}^{T_0} 
- \oint dt \phi {\dot E}$, we keep only the change in energy, and neglect the change in the turning point to this order.  Using the result (\ref{edot}), these approximations imply that $\oint d\phi\delta E = \Phi_0^2 G_S \oint dt (\phi_0 - \phi) {\dot \phi}^2$.  Collecting results, we find
\begin{equation}
\langle {\dot \phi}^3\rangle_E  = \frac{2\Phi_0^2 G_S}{m}  
\langle (\phi_0 - \phi) {\dot \phi}^2\rangle_E.
\label{third1}
\end{equation}
Recalling the definition $m = \Phi_0^2 C$, and inserting (\ref{third1}) 
into (\ref{result}), we recover our main result (\ref{WDR}). 

\bibliographystyle{apsrev}

\begin{thebibliography}{99}

\bibitem{Reulet}
B. Reulet, J. Senzier, and D. E. Prober, Phys. Rev. Lett. {\bf 91}, 196601 (2003).

\bibitem{Levitov1}
L. S. Levitov, and M. Reznikov,
Phys. Rev. B {\bf 70}, 115305 (2004).

\bibitem{Nagaev}
K. E. Nagaev, Phys. Rev. B {\bf 66}, 075334 (2002).

\bibitem{Beenakker}
C. W. J. Beenakker, M. Kindermann, and Yu. V. Nazarov, Phys. Rev. Lett. {\bf 90}, 176802 (2003). 

\bibitem{Bomze}
Yu. Bomze {\it et al.}, Phys. Rev. Lett. {\bf 95}, 176601 (2005).

\bibitem{Simon}
S. Gustavsson {\it et al.}, Phys. Rev. Lett. {\bf 96}, 076605 (2006).

\bibitem{Nazarov}
J. Tobiska and Yu. V. Nazarov, Phys. Rev. Lett. {\bf 93}, 106801 (2004).

\bibitem{ourPRB} 
A. N. Jordan and E. V. Sukhorukov, Phys. Rev. B {\bf 72}, 035335 (2005).

\bibitem{Kramers}
H. A. Kramers, Physica (Utrecht) {\bf 7}, 284 (1940).

\bibitem{ourprl} 
A. N. Jordan and E. V. Sukhorukov, Phys. Rev. Lett. {\bf 93}, 260604 (2004).

\bibitem{Pekola}
J.P. Pekola {\it et al.}, Phys. Rev. Lett. {\bf 95}, 197004 (2005).

\bibitem{Ankerhold}
J. Ankerhold, cond-mat/0607020.

\bibitem{Pekola2}
A.V. Timofeev {\it et al.}, cond-mat/0612087.

\bibitem{Saclay}
H. Pothier, B. Huard, N. Birge, D. Esteve, unpublished.

\bibitem{SPI}
S. Pilgram, A. N. Jordan, E. V. Sukhorukov, and M. B\"uttiker, Phys. Rev. Lett. {\bf 90}, 206801 (2003). 

\bibitem{JMP} 
A. N. Jordan, E. V. Sukhorukov, and S. Pilgram, J. Math. Phys. {\bf 45}, 4386 (2004). 

\bibitem{footnote0}
Regarding circuit effects, our results do not appear to coincide with
Ref.\ [\onlinecite{Ankerhold}].  

\bibitem{book} 
C. W. Gardiner, {\it Handbook of Stochastic Methods} (Springer-Verlag, Berlin, 1990).


\bibitem{footnote1} The prefactor in $\Gamma$ remains unknown, 
because it depends on non-Markovian behavior at short times 
of order $\tau_0$. For a discussion of the prefactor
in a JJ context, see, e.g., M. B\"uttiker, E. P. Harris, and 
R. Landauer, Phys. Rev. B {\bf 28}, 1268 (1983). 

\bibitem{footnote2}
The canonical ``time'' variable $s$ differs in general from the physical time $t$.

\bibitem{footnote3}
The contribution $\langle\dot\phi\ddot\phi\rangle_E$,
which arises from retardation of the average current
response to the slowly varying voltage with $\dot V=\Phi_0\ddot\phi$,
may be taken into account by rescaling the capacitance $C$.

\bibitem{footnote4}
E. V. Sukhorukov, unpublished. 

\bibitem{amplifier}
I. Siddiqi {\it et al.}, Phys. Rev. Lett. {\bf 93}, 207002 (2004).

\end{thebibliography}

\end{document}